\documentclass[11pt]{article}
\pdfoutput=1
\usepackage{jheppub,amsmath,amssymb,hyperref}

\usepackage{graphicx}
\usepackage{todonotes}
\usepackage[mathscr]{eucal}
\usepackage{amsmath}
\usepackage{color}
\usepackage[utf8]{inputenc}

\newcommand{\gsimm}{\raise.3ex\hbox{$>$\kern-.75em\lower1ex\hbox{$\sim$}}}
\newcommand{\lsimm}{\raise.3ex\hbox{$<$\kern-.75em\lower1ex\hbox{$\sim$}}}

\newcommand{\comment}[1]{}

\newcommand{\be}{\begin{equation}}
\newcommand{\ee}{\end{equation}}
\newcommand{\ba}{\begin{eqnarray}}
\newcommand{\ea}{\end{eqnarray}}

\newcommand{\bea}{\begin{eqnarray*}}
\newcommand{\eea}{\end{eqnarray*}}

\def\b{\beta}

\begin{document}
\title{Gravitational Waves in  Doubly Coupled Bigravity}

\author[a]{Philippe Brax,}
\author[b]{Anne-Christine Davis,}
\author[c]{Johannes Noller}

\affiliation[a]{Institut de Physique Th\'{e}orique, Universit\'e Paris-Saclay, CEA, CNRS, F-91191 Gif/Yvette Cedex, France}
\affiliation[b]{DAMTP, Centre for Mathematical Sciences, University of Cambridge, CB3 0WA, UK}
\affiliation[c]{Astrophysics, University of Oxford, DWB, Keble Road, Oxford, OX1 3RH, UK}

\emailAdd{philippe.brax@cea.fr}
\emailAdd{A.C.Davis@damtp.cam.ac.uk}
\emailAdd{noller@physics.ox.ac.uk}

\date{today}
\abstract{We consider gravitational waves from the point of view of both their production and their propagation in doubly coupled bigravity in the metric formalism. In bigravity, the two gravitons are coupled
by a non-diagonal mass matrix and show birefrigence. In particular, we find that one of the two gravitons propagates with a speed which differs from one. This deviation is tightly constrained by both the gravitational Cerenkov effect and the energy loss of binary pulsars. When emitted from astrophysical sources, the Jordan frame gravitational wave, which is a linear combination of the two propagating gravitons, has a wave form displaying beats. The best prospect of detecting this phenomenon would come from nano-Hertz interferometric experiments.  }



\maketitle

\setcounter{tocdepth}{2}

\section{Introduction}
The recent direct detection of gravitational waves  \cite{Abbott:2016blz,Abbott:2016nmj} as predicted by General Relativity (GR) \cite{Buonanno:2007yg}  one hundred years ago could also serve as a test for alternative theories of gravity.  For instance  a loose bound on the deviation of the speed of gravitational waves from the speed of light has been extracted from the recent LIGO events \cite{Blas:2016qmn}. Hence  gravitational waves can be used to constrain certain modified gravity theories. Motivated by the late time acceleration of the expansion of the Universe \cite{Copeland:2006wr,Joyce:2014kja}, models of massive gravity \cite{deRham:2010ik,deRham:2010kj} have been recently considered where gravity could be the result of the existence of two or more gravitons \cite{Hassan:2011hr,Hassan:2011zd}. In the case of bigravity, { the general case we will consider here} is that of doubly coupled models whereby a linear combination of the two gravitons couple to matter \cite{deRham:2014naa,Noller:2014sta}. The { gravitational wave} phenomenology of the singly coupled case has already been considered \cite{DeFelice:2013nba,Narikawa:2014fua} with the existence of beats in the wave form, which could be detectable by LIGO only if the speed of gravitational waves is extremely close to one. In this paper, we generalise these results to the doubly coupled case, where the amplitude and the phase of the Jordan frame wave is shown to have differing characteristics from the singly coupled case. For instance, the modulation of the GR wave emitted by far away sources does not vanish at large frequency any more.

In bigravity, the two gravitons obey coupled propagation equations with eigenmodes whose speeds deviate from one. In this paper, we focus on the cosmological models where the graviton mass is of order of the Hubble rate now - { the background and perturbative cosmology of such doubly coupled models has previously been explored in} \cite{Enander:2014xga,Comelli:2015pua,Gumrukcuoglu:2015nua,Lagos:2015sya}.
On scales much shorter than the size of the Universe, the mass terms can be neglected and the emission from local sources resembles the one in GR for each individual graviton. We examine the emission from such sources and apply it to the case of binary pulsars. The energy loss is modified compared to GR, which results in a tight bound on the deviation of the speed of gravitational waves at the per mil level \cite{Jimenez:2015bwa}. Once emitted and far away from the source, these waves propagate like plane waves which mix and show birefringence, i.e. the Jordan frame gravitational wave can be expressed as an effective propagation wave with a frequency dependent amplitude and phase shift whilst the effective  gravitational speed differs from one and is also frequency dependent. The gravitational Cerenkov effect when the effective speed is smaller than the speed of light leads to an even tighter bound \cite{Moore:2001bv,Kimura:2011qn,Kimura:2016voh} than the one from binary pulsars.

In view of the recent direct detection of gravitational waves, one may enquire whether gravitational birefringence could be observed. This would require to disentangle the frequency dependence of the wave form from its amplitude, as the amplitude would be degenerate with the features, such as the masses, of the emitting system. We find that this can only be envisaged at best in the nano-Hertz regime \cite{Manchester:2010tp} and for small differences between the effective gravitational speed and the speed of light. Otherwise, it is likely that the modulation of the bigravity signal would be averaged out resulting in an undetectable change of the wave amplitude.

The paper is arranged as follows. In section 2, we recall the main features of doubly coupled bigravity. In section 3, we consider the tensor modes and their emission from local sources. This allows us to use the binary pulsars to put a bound on the effective speed of gravity. In section 4, we analyse the propagation from a distant source and in section 5 the prospect of detecting the effects of gravitational birefringence.

\section{Bigravity}
\subsection{The model}


We consider massive bigravity models coupled to matter in the constrained vielbein formalism, which is equivalent to the metric formulation \cite{Brax:2016ssf}, for energy scales below the strong coupling limit $\Lambda_3\sim (M_{\rm Pl} H_0^2)^{1/3}$ corresponding to scales larger than 1000 km's.\footnote{Technically speaking this is the scale where perturbative unitarity is lost for fluctuations around Minkowski. While this is therefore an excellent guess for the cutoff scale, whether full unitarity is lost at $\Lambda_3$, i.e. whether this scale is a strict cutoff, is still not known. Also note that, for backgrounds different to Minkowski, this scale will get re-dressed. For example ratios of the scale factors in the theory will modify this scale, when FRW backgrounds are chosen for both metrics.}  Bigravity can be formulated  using two vielbeins $e_{1\mu}^a$ and $e_{2\mu}^a$ \cite{Hinterbichler:2012cn}, which couple to matter with couplings $\beta_{1,2}$ respectively \cite{deRham:2014naa,Noller:2014sta}.\footnote{ Note that in general other consistent non-derivative matter couplings exist \cite{Melville:2015dba}, but when enforcing the symmetric vielbein condition (as we do here) the couplings of \cite{deRham:2014naa,Noller:2014sta} are the unique consistent matter couplings \cite{Melville:2015dba,deRham:2015cha,Matas:2015qxa,Heisenberg:2015iqa,Huang:2015yga}. In this context also note the derivative couplings of \cite{Heisenberg:2015wja}.} The action comprises three very distinct parts. The first one is simply the Einstein-Hilbert terms for both metrics $g^{1,2}_{\mu\nu}$ built from the two vielbeins
\be
S_G= \int d^4x\ e_1 \frac{R_1}{16\pi G_N} + \int d^4 x\ e_2 \frac{R_2}{16\pi G_N}
\ee
where $R_{1,2}$ are the Ricci scalars built from the respective metrics, and $e_{1,2}$ are the determinants of the vielbeins viewed as $4\times 4 $ matrices.
The individual vielbeins $e^a_{\alpha\mu}$, $\alpha=1,2$ are constrained to satisfy the symmetric condition
\be
e_{1\mu}^a e_{2\nu}^b \eta_{ab}=e_{1\nu}^a e_{2\mu}^b \eta_{ab},
\label{sym}
\ee
which { we explicitly enforce}.
This ensures the equivalence with doubly coupled bigravity in the metric formulation, in particular all the terms in the action can be written in terms of the two individual metrics $g^\alpha_{\mu\nu}, \ \alpha=1,2$
defined by
\be
g^{\alpha}_{\mu\nu}= \eta_{ab} e^a_{\alpha\mu}e^{b}_{\alpha \nu}.
\ee
Matter, i.e. all the fields of the standard model of particle physics, couple to the Jordan metric
 \be
g_{\mu\nu}= \eta_{ab} e^a_\mu e^b_\nu
\ee
built from
the local frame \cite{Noller:2014sta}
\be
e^a_\mu = \beta_1 e_{1\mu}^a+\beta_2 e_{1\mu}^a
\ee
where $a$ is a local Lorentz index and $\mu$ the global coordinate index associated with the one forms $e^a= e^a_\mu dx^\mu$.
The Jordan metric $g_{\mu\nu}$
is explicitly related to the $g^\alpha_{\mu\nu}$'s by
\be
g_{\mu\nu}= \beta_1^2 g^1_{\mu\nu} + \beta_1 \beta_2 Y_{\mu\nu} + \beta_2^2 g^2_{\mu\nu}
\ee
where we have defined the symmetric tensor
\be
Y_{\mu\nu}= \eta_{ab}( e^a_{1\mu} e^b_{2 \nu}+ e^a_{2\mu} e^b_{1 \nu})
\ee
which is also directly linked to $g^\alpha_{\mu\nu},\ \alpha=1,2$ as the symmetric condition is enforced.

Matter fields $\psi_i$ are (minimally) coupled to $g_{\mu\nu}$ and  the matter action involves the coupling of the matter fields $\psi_i$'s to the Jordan metric $g_{\mu\nu}$
\be
S_m(\psi_i, g_{\mu\nu}).
\ee
Massive bigravity involves also a potential term \cite{Hinterbichler:2012cn,Hassan:2011hr, Hassan:2011zd}
\be
S_V=\Lambda^4 \sum_{ijkl} m^{ijkl} \int d^4 x\  \epsilon_{abcd} \epsilon^{\mu\nu\rho\sigma} e^a_{i\mu} e^b_{j\nu}e^c_{k\rho} e^d_{l\sigma}
\ee
where
\be \Lambda^4= m^2 M^2_{\rm Pl}
\ee
and $m$ is related to the graviton mass while the dimensionless and fully symmetric tensor $m^{ijkl}$ involves five real coupling constants of order one.
Both the matter coupling and the potential terms can be expressed as a function of the individual metrics $g^\alpha_{\mu\nu}$.

The Jordan frame energy-momentum tensor is defined by
\be
T_{\mu\nu}= -\frac{2}{e} \frac{\delta S_m}{\delta g^{\mu\nu}},
\ee
which is obtained by varying the matter action with respect to the Jordan metric, i.e. not with respect to the two metrics $g^\alpha_{\mu\nu}$. The Einstein equations for both metrics which follow from this setting read
\be
G^{1}_{\mu\nu} = 8\pi G_N ( T^{1}_{\mu\nu} + {\cal T}^{1}_{\mu\nu})
\ee
and
\be
G^{2}_{\mu\nu} = 8\pi G_N ( T^{2}_{\mu\nu} + {\cal T}^{2}_{\mu\nu})
\ee
where we have introduced the tensors
\be
T^{\alpha}_{\mu\nu}= -\frac{2}{e_\alpha} \frac{\delta S_m}{\delta g_\alpha^{\mu\nu}},\ \ {\cal T}^{\alpha}_{\mu\nu}= -\frac{2}{e_\alpha} \frac{\delta S_V}{\delta g_\alpha^{\mu\nu}}
\ee
from which both the background cosmology and the gravitational wave equations can be deduced. In the following, we will recall how the background cosmological solutions appear. For gravitational waves, we will derive them by directly using the Lagrangian of bigravity at the second order level in the gravitational perturbations.

\subsection{Cosmological background}

The previous model  can be specialised by choosing the cosmological ansatz for the metrics
\be
ds_1^2= a_1^2 (-d\eta^2 +dx^2)
\ee
and
\be
ds_2^2= a_2^2 (- b^2 d\eta^2 +dx^2)
\ee
where the ratio between the lapse functions $b^2$ plays a crucial role in the modification of gravity induced by the bigravity models.
We consider the coupling of bigravity to a perfect fluid defined by the energy-momentum tensor
\be
T^{\mu\nu}= (\rho+p) u^\mu u^\nu + p g^{\mu\nu}
\ee
where the 4-vector $u^\mu$ is
$
u^\mu= \frac{dx^\mu}{d\tau_J}
$
and the proper time in the Jordan frame is simply
$
d\tau_J^2 =- g_{\mu\nu}dx^\mu dx^\nu.
$
Using the fact that the Jordan interval is given by
\be
ds^2= -(\beta_1 a_1 + \beta_2 b a_2)^2 d\eta^2 + (\beta_1 a_1 +\beta_2 a_2)^2 dx^2
\ee
we can identify the Jordan frame scale factor
\be
a_J= \beta_1 a_1 +\beta_2 a_2
\ee
and the conformal times
\be
d\eta_1 =d\eta,\ \ d\eta_2= b d\eta
\ee
when  the Jordan conformal time is
\be
d\eta_J= \frac{\beta_1 a_1 + \beta_2 b a_2}{\beta_1 a_1 +\beta_2 a_2}d\eta.
\ee
Matter is conserved in the Jordan frame, as follows from the residual diffeomorphism invariance  of the matter action,  implying that
\be
\frac{d\rho}{d\eta_J} + 3a_J { H}_J (\rho+p)=0
\ee
where the Jordan frame Hubble rate is identified with
\be
H_J\equiv \frac{d a_J}{a_J^2 d\eta_J}= \frac{1}{(\beta_1 a_1 + \beta_2 b a_2) a_J}(\beta_1 a_1^2 H_1  + \beta_2 a_2^2 H_2)
\ee
and we have introduced the two Hubble rates
$
H_1=\frac{d a_1}{a_1^2 d\eta_1}\equiv \frac{d a_1}{a_1^2 d\eta},\ H_2=\frac{d a_2}{a_2^2 d\eta}.
$
The cosmological dynamics are governed by the two  Friedmann equations
\be
3H_1^2 M_{\rm Pl}^2=  \beta_1 \frac{a_J^3}{a_1^3}\rho +24\Lambda^4 m^{1jkl}\frac{a_j a_k a_l}{a_1^3}.
\label{F1}
\ee
and
\be
\frac{3H_2^2 M_{\rm Pl}^2}{b^2}=  \beta_2 \frac{a_J^3}{a_2^3}\rho +24\Lambda^4 m^{2jkl}\frac{a_j a_k a_l}{a_2^3}.
\label{F2}
\ee
These equations have two types of solutions. Here
we consider only the branch of solutions which satisfies the constraint
\be
b=\frac{a_2 H_2}{a_1 H_1}.
\label{bi}
\ee
It turns out that the dynamics simplify both at late and early times.
 When dark energy is negligible, i.e. in the radiation and matter eras, we have that the ratio $X= \frac{a_2}{a_1}$ converges to a constant
\be
X\to X_m= \frac{\beta_2}{\beta_1}
\ee
and in the asymptotic future when dark energy dominates, i.e. when the terms in $\Lambda^4$ in both Friedmann equations (\ref{F1}) and (\ref{F2}) are dominant,  we have that
\be
X\to X_d
\ee
 where
 \be
X_d= \frac{m^{2jkl} a_j a_k a_l}{m^{1jkl} a_j a_k a_l}.
\ee
In both cases we have that
\be
b=1.
\ee
\begin{figure*}
\centering
\includegraphics[width=0.59\linewidth]{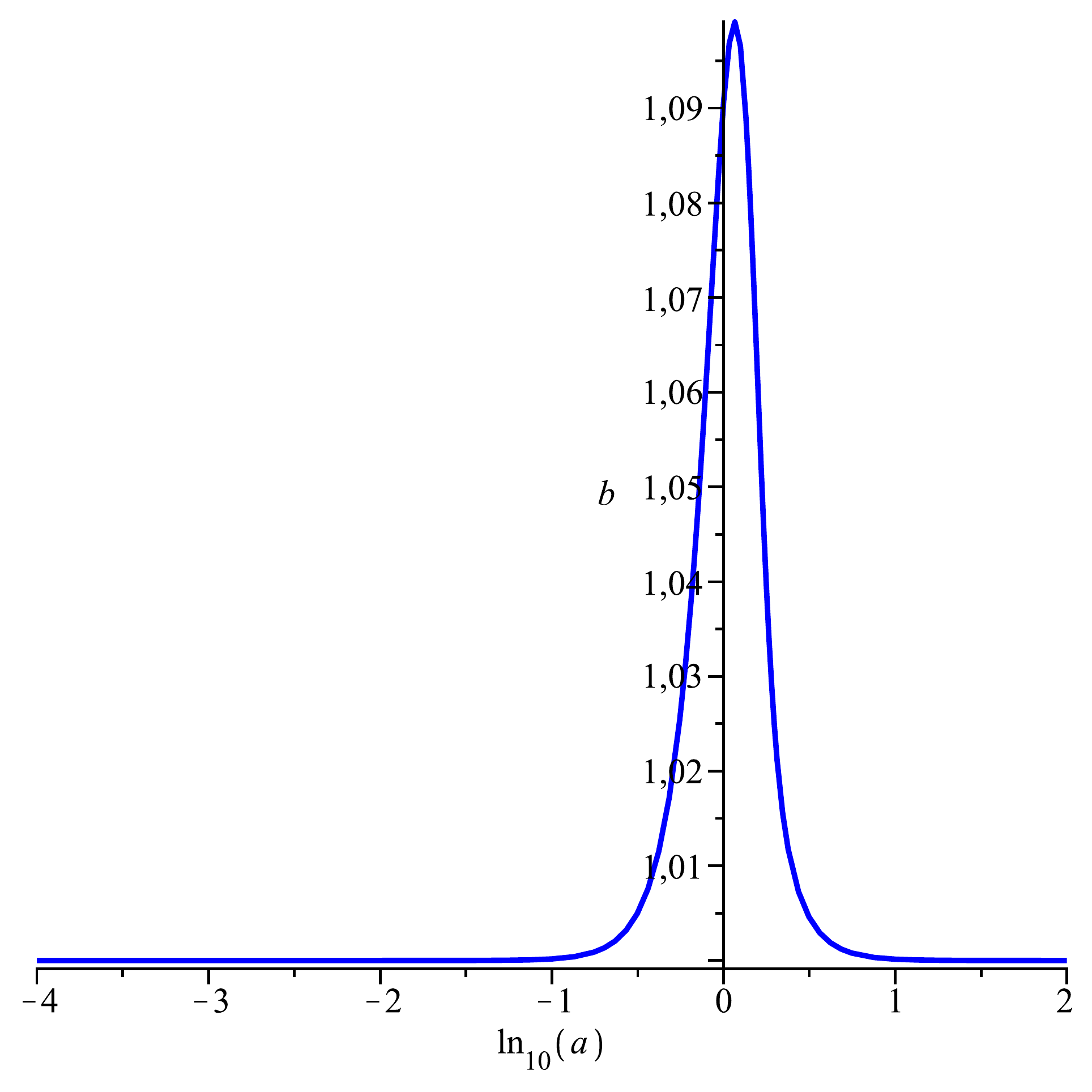}
\caption{The variation of $b$ as a function of $\ln a_J$ for a model where all the $m_{ijkl}=1$, $\beta_1=1.1$ and $\beta_2=1$. In the recent past of the Universe, $b$ starts deviating from 1 before settling back to one in the far future.}
\end{figure*}
Between these eras, and in particular now, $b\ne 1$ and $X$ is not equal to its asymptotic value \cite{Brax:2016ssf}, see figure 1. This will prove to be particularly important for gravitational waves as the effective speed of propagation deviates from one when $b\ne 1$, i.e. we can expect to have non-standard gravitational wave propagation in the recent Universe.

\section{Tensor modes: emission and propagation}
\subsection{Propagation equations}

There are two gravitons in bigravity models. They can be characterised using
the tensor perturbations of the two vielbeins
\be
\delta e^{\alpha i}_j= a_\alpha h^{i}_{\alpha j}
\ee
where $\alpha=1,2$ and $h^{i}_{\alpha j}$ is a symmetric transverse and traceless tensor with two degrees of freedom. In the rest of this paper, we do not consider scalar and vector perturbations and only concentrate on the helicity two parts of the perturbations \cite{Brax:2016ssf}. The potential term of bigravity induces a mass term for the gravitons  which reads
\be
M^2_{\alpha\beta}(a_\gamma)= -24 m^2 (b_\alpha b_\beta)^{1/2}m_{\alpha\beta}(a_\gamma)
\label{mass}
\ee
which is a symmetric matrix of order $m^2$
where
\be
m_{\alpha\beta}(a_\gamma)= \sum_{\gamma\delta} m_{\alpha\beta\gamma\delta}\tilde a_\gamma a_\delta.
\ee
and $\tilde a_\alpha= b_{\alpha}a_{\alpha}$ with $b_1=1$ and $b_2=b$.
We have normalised the tensor modes according to
\be
 \bar h^1_{ij}=  M_{\rm Pl} a_1 h^1_{ij},\ \bar h^2_{ij}=  M_{\rm Pl} \frac{a_2}{b^{1/2}} h^2_{ij}.
\ee
Notice that the mass matrix is not diagonal and evolves with time. This induces a mixing of the two gravitons, i.e. birefrigence.
The evolution equations for the two gravitons $h_1$ and $h_2$  can be deduced from the action expanded to second order in the perturbations and read
\be
\frac{d^2{\bar {h }}_1}{d\eta^2}-\Delta \bar h_1 +(M^2_{11}(a_\gamma)- \frac{1}{a_1} \frac{d^2 a_1}{d\eta^2}) \bar h_1 + M^2_{12}(a_\gamma) \bar h_2=0
\label{pro1}
\ee
and
\be
\frac{d^2 {\bar {h }}_2}{d\eta^2}-b^2 \Delta \bar h_2 + (M^2_{22}(a_\gamma)-\frac{b^{1/2}}{a_2} \frac{d^2 (a_2b^{-1/2})}{d\eta^2}) \bar {h_2} + M^2_{21}(a_\gamma) \bar h_1=0.
\label{pro2}
\ee
The coupling between the two gravitons will induce beats in the Jordan gravitational waves. This follows from the fact that matter couples to the Jordan frame combination of gravitons
\be
a_J h^i_{jJ}= \beta_1 a_1 h^{i}_{j1}+ \beta_2 a_2 h^{i}_{j2}
\ee
and one can see that this evolves with time, i.e. matter couples to different gravitons in the history of the Universe.

\subsection{Gravitational waves from local sources}

Let us now consider a gravitational source and the way gravitational waves are emitted. This can be conveniently analysed starting from the action of the two gravitons coupled to matter. Let us recall first how this operates in General Relativity.  The action involves
\be
{\cal L}_{\rm GR}= \frac{1}{2} ( \frac{d\bar h_{ij}}{d\eta}\frac{d\bar h^{ij}}{d\eta} -\vec\nabla \bar h_{ij}\vec \nabla \bar h^{ij}+ \frac{1}{a} \frac{d^2 a}{d \eta^2} \bar h^{ij} \bar h_{ij})
+ \frac{a}{M_{\rm Pl}}  \bar h^{ij} \bar T_{ij}
\ee
where $\bar T_{ij}=T_{ij}-\frac{\delta_{ij}}{3}T$ and indices are raised with $\delta^{ij}$.
The gravitational equation becomes
\be
\frac{d^2{\bar {h_{ij} }}}{d\eta^2}-\Delta \bar h_{ij}- \frac{1}{a} \frac{d^2 a}{d\eta^2} \bar h_{ij} = \frac{a}{M_{\rm Pl}}\bar T_{ij}
\label{pro3}
\ee
where here $a$ is the scale factor of the FRW Universe and $\bar T_{ij}$ the traceless part of the spatial energy momentum tensor. Notice that in General Relativity we have $8\pi G_N= M_{\rm Pl}^{-2}$.
In bigravity,
 matter couples to the Jordan frame energy-momentum tensor too via
\be
S_{\rm in}= \int d^4x \tilde a_J  (\beta_1 \bar h_1^{ij } + \beta_2 b^{1/2} \bar h_2^{ij}) \bar T_{ij}.
\ee
As a result the coupled gravitational equations become
\be
\frac{d^2{\bar h_{ij}^1}}{d\eta^2}-\Delta \bar h^1_{ij} +(M^2_{11}(a_\gamma)- \frac{1}{a_1} \frac{d^2 a_1}{d\eta^2}) \bar h^1_{ij} + M^2_{12}(a_\gamma) \bar h^2_{ij}= \beta_1 \frac{\tilde a_J}{M_{\rm Pl}}\bar T_{ij}
\label{pro1}
\ee
and
\be
\frac{d^2 {\bar {h }}^2_{ij}}{d\eta^2}-b^2 \Delta \bar h^2_{ij} + ((M^2_{22}(a_\gamma)-\frac{b^{1/2}}{a_2} \frac{d^2 (a_2b^{-1/2})}{d\eta^2}) \bar h^2_{ij} + M^2_{21}(a_\gamma) \bar h^1_{ij}= \beta_2 \frac{b^{1/2} \tilde a_J}{M_{\rm Pl}}\bar T_{ij}.
\label{pro2}
\ee
In the following, we shall
be only interested in waves which propagate on distances for which one can neglect the effects of the cosmological evolution. The generalisation to the cosmological case can be easily analysed too and is left for future work.  We will also assume that the waves are emitted at a redshift corresponding to  $a_J$ in bigravity and $ a_{\rm GR}$   in $\Lambda$-CDM. Both in bigravity and in GR, the scale factors $a_J$ and $a_{\rm GR}$ are  normalised to be one now. As a result, the metrics read
\be
ds^2_{\rm GR}\sim -dt_{\rm GR}^2 + d\vec r_{\rm GR}^2
\ee
and
\be
ds^2_J\sim -dt^2_J + d\vec r_{J}^2
\ee
where $dt_J= \tilde a_J d\eta$ and $dt_{\rm GR} = a_{\rm GR} d\eta$. Moreover we have $d\vec r_J= a_J d\vec x$ and $ d\vec r_{\rm GR}= a_{\rm GR} d\vec x$. As $a_{\rm GR} \sim a_J \sim 1$ in the recent past of the Universe, the only difference between the two metrics now comes from the different clocks with $a_J= \beta_1 a_1 + \beta_2 a_2$ and $\tilde a_J= \beta_1 a_1 + \beta_2 b a_2$ when $b\ne 1$. We also assume that the waves can be well approximated by plane waves sufficiently far from the source.

\subsection{Emission from binary pulsars}

The emission of gravitational waves by binary pulsars leads to tight constraints on modified gravity. Here the emission takes places on scales much smaller than the inverse mass of the gravitons, i.e. less than the size of the Universe. The perturbative equations that we adopt are only valid at low energy corresponding to time scales larger than the inverse cut-off $\Lambda_3^{-1}\sim 10^{-2}$ s. As the typical period of binary pulsars is of the order of a few hours, the description which follows, where the  emission of gravitational waves is considered in bigravity, can be applied to binary pulsars.  The wave equations in the emission region therefore simplify
\be
\frac{d^2\bar h_{ij}^1}{d\eta^2}-\Delta \bar h^1_{ij} = \beta_1 \frac{\tilde a_J}{M_{\rm Pl}}\bar T_{ij}
\label{pro1}
\ee
and
\be
\frac{d^2 {\bar {h }}^2_{ij}}{d\eta^2}-b^2 \Delta \bar h^2_{ij} = \beta_2 \frac{\tilde a_J b^{1/2}}{M_{\rm Pl}}\bar T_{ij}.
\label{pro2}
\ee
{The Newtonian trajectories of the binary objects are not modified in doubly-coupled bigravity (see section 5 of \cite{Brax:2016ssf}) and here we consider that this is still a reasonable assumption in the case of compact objects with Newtonian potentials $\Phi_N \lesssim 0.1$.}
The solutions to the wave equations are simply
\be
\bar h_{ij}^1(\vec x,\eta)= \beta_1 \frac{\tilde a_J}{4\pi M_{\rm Pl}}\Lambda_{ij}^{kl} \int d^3 y \frac{\bar T_{kl} (\eta -\vert x- y\vert, y)}{\vert x-y\vert}
\ee
in conformal coordinates and
\be
\bar h_{ij}^2(\vec x,\eta)= \beta_2 \frac{\tilde a_J b^{1/2}}{4\pi M_{\rm Pl}} \Lambda_{ij}^{kl} \int d^3 y \frac{\bar T_{ij} (\eta -\frac{\vert x- y\vert}{b}, y)}{\vert x-y\vert}.
\ee
Assuming that the energy-momentum tensor of the source has compact support and $\vert x\vert \gg \vert y\vert$ we have the approximation
\be
\bar h_{ij}^1(\vec x,\eta)= \beta_1 \frac{\tilde a_J}{4\pi M_{\rm Pl}} \frac{1}{\vert x\vert}\Lambda_{ij}^{kl}\int d^3 y {\bar T_{ij} (\eta -\vert x\vert, y)}
\ee
in conformal coordinates and
\be
\bar h_{ij}^2(\vec x,\eta)= \beta_2 \frac{\tilde a_J b^{1/2}}{4\pi M_{\rm Pl}}\frac{1}{\vert x\vert} \Lambda_{ij}^{kl}\int d^3 y {\bar T_{kl} (\eta -\frac{\vert x\vert}{b}, y)}.
\ee
Using the identity
\be
\frac{d^2}{d\eta^2} \int d^3y y^i y^j  T^{00}(\eta -\vert x\vert, y)= 2 \int d^3 y T^{ij} (\eta -\vert x\vert,y)
\ee
for the conserved energy-momentum in the Jordan frame, we find that
\be
\bar h_\alpha^{ij} (\vec x,\eta)= \beta_\alpha b_\alpha^{1/2} \frac{a_J^4}{\tilde a_J 8\pi M_{\rm Pl}}\frac{1}{ \vert x \vert}  \Lambda^{ij}_{kl}\frac{d^2\bar I^{kl}}{d\eta^2}
\ee
where indices
are raised and lowered with  the flat $\delta^{ij}$. This implies that
\be
\bar h_\alpha^{ij} (\vec x,\eta)=\frac{\beta_\alpha b_\alpha^{1/2}}{\beta_1^2+\beta_2^2} \frac{a_J^4}{\tilde a_J a_{\rm GR}^3} \bar h_{\rm GR}^{ij}(\vec x,\eta).
\label{ini}
\ee
where we have used the fact that the local and cosmological Newton constants in bigravity models is \cite{Brax:2016ssf}
\be
G_{\rm local}=G_{\rm cosmo}= (\beta_1^2 +\beta_2^2 )G_N
\ee
where $G_N$ is only a parameter in the action.
We have introduced the usual tensor
\be
\Lambda_{ij}^{kl}=P^k_iP^j_l- \frac{1}{2} P_{ij}P^{kl}
\ee
where
\be
P_{ij}= \delta_{ij}-n_in_j
\ee
which is the projector orthogonal to the propagation vector $n_i$. The tensor $\Lambda_{ij}^{kl}$ enforces the transverse traceless condition.
We also have the Jordan combination
\be
h^J_{ij} (\vec x,\eta)= G_{\rm cosmo}\frac{ a_J^3}{\tilde a_J}\frac{\beta_1^2 + b \beta_2^2}{\beta_1^2+\beta_2^2}\frac{1}{ \vert x \vert} \Lambda_{ijkl}\frac{d^2\bar I^{kl}}{d\eta^2}
\ee
where, in terms of the matter density $\rho$,
\be
\bar I^{ij}= \int d^3 y (y^iy^j-\frac{1}{3} \delta^{ij} \vert y\vert ^2) \rho(\eta-\vert x\vert, y).
\ee
to leading order in a multipolar expansion. We have assumed that $b$ is very close to unity.

The energy flux emitted by the object can be evaluated as in \cite{Schutz} where it is the energy given to matter minus the one that matter radiates subsequently. As the gravitational waves couple to matter in the Jordan frame, this depends only on the derivatives of $h_J$
\be
F= \frac{1}{8 \pi a_J^4} < (\frac{d h_{ij}}{dt_J})^2>
\ee
where the average is a time average.
The energy loss is given
\be
\frac{dE}{dt_J}= - \int F a_J^2 \vert x\vert^2  d\Omega
\ee
and therefore
\be
\frac{dE}{dt_J}= - \frac{a_J^4}{2\tilde a_J^4} G^2_{\rm cosmo} (\frac{\beta_1^2 +b \beta_2^2}{\beta_1^2 +\beta_2^2})^2 < (\dddot {\cal I}_{ij})^2>
\ee
where the time derivatives are with respect to $\eta$.  As a result
\be
\frac{dE}{dt_J}=(\frac{\beta_1^2 +b \beta_2^2}{\beta_1^2 +\beta_2^2})^2 \frac{a_J^4}{\tilde a_J^4}\frac{dE}{dt_{\rm GR}}.
\ee
Notice that the deviation from the GR result is only present when $b\ne 1$. As we have already recalled, this is the case in the present Universe. There is a tight constraint on the possible difference with GR and it reads \cite{Jimenez:2015bwa}
\be
0.995<(\frac{\beta_1^2 +b \beta_2^2}{\beta_1^2 +\beta_2^2})^2 \frac{a_J^4}{\tilde a_J^4}<1
\ee
which gives a constraint on $b$ at the $10^{-3}$ level. In the following, we shall investigate what happens  to the propagation of the gravitational waves when $b$ is constrained at a level tighter than one per mil.

Our calculation has taken into account the quadrupolar emission from binary pulsars. In this case, the distance between the two stars is much larger than the cut-off distance of bigravity and our calculation is valid where the two stars are considered to be orbiting subject to Newton's law.

{On the other hand, since the stars themselves (typically neutron stars) are much smaller than the cut-off scale of bigravity, their dynamics will most likely be sensitive to details of the UV completion of the theory. For example, the additional decoupled scalar degree of freedom of doubly coupled bigravity \cite{Brax:2016ssf}, which naively becomes a ghost below the cut-off distance, may correspond to a healthy degree of freedom in the UV-completed theory and lead to stars acquiring scalar charges.} This would lead to the possible emission of dipolar gravitational waves \cite{Damour:1992we,Yagi:2013ava}. Another phenomenon which is beyond the present treatment corresponds to the last phase of the merger between two black holes when their distance falls below 1000 km's. The calculation of the emission spectrum cannot be tackled using the models described here. All these effects are beyond the present work.

\section{Propagation}

Let us come back to the propagation of gravitational waves in empty space, when the initial wave is due to a localised source which is far-away and the waves can be considered to be plane-waves.
\subsection{Eigenmodes}

It is convenient to define the
 effective mass matrix
\be
\tilde M^2=  \left (
\begin{array}{cc}
M_{11}^2- \frac{1}{a_1} \frac{d^2 a_1}{d\eta^2}&M_{12}^2 \\
M_{12}^2 &M_{22}^2 -\frac{b^{1/2}}{a_2} \frac{d^2 (a_2b^{-1/2})}{d\eta^2}) \\
\end{array}
\right ).
\ee
The two propagation equations for gravitons have
 two eigenmodes which can be described by
\be
h_{\alpha\pm}= A_{\alpha\pm} e^{i(\omega_\pm t- i \vec k.\vec x)}
\ee
where $\alpha=1,2$. The eigenfrequencies are given by the quartic dispersion relation
\be
\omega^4 - \omega^2 ( (1+b^2)\vec k^2 + \tilde M_{11}^2 +\tilde M_{22}^2) - \tilde M_{12}^2 \tilde M^2_{21} +  (\vec k^2 + \tilde M^2_{11})(\b^2\vec k^2 + \tilde M_{22}^2)=0.
\ee
Defining the discriminant
\be
\Delta= ( (1-b^2)\vec k^2 + \tilde M_{11}^2 -\tilde M_{22}^2)^2 +4 \tilde M_{12}^2 \tilde M^2_{21}
\ee
we have the two eigenfrequencies
\be
\omega^2_{\pm}= \frac{\omega^2 ( \vec k^2 + \tilde M_{11}^2 +b^2 \vec k^2 +\tilde M_{22}^2)\pm \sqrt{\Delta}}{2}.
\ee
We only consider gravitational waves such that $\vec k^2 \gg \tilde M^2_{ij}$ as the mass matrix elements are of order $H_0$ and astrophysical waves are much more energetic than this. As a result we obtain the expansion
\ba
&&\omega_+^2\sim \vec k^2 +\tilde M_{11}^2 + \frac{\tilde M_{12}^4}{(1-b^2)\vec k^2 +\tilde M_{11}^2-\tilde M^2_{22}}\nonumber \\
&&\omega_-^2\sim  b^2 \vec k^2  +\tilde M_{22}^2 - \frac{\tilde M_{12}^4}{(1-b^2)\vec k^2 +\tilde M_{11}^2-\tilde M^2_{22}}.\nonumber \\
\ea
The two eigenmodes are then obtained as
\be
h_-= h_2 -C h_1, \ \ h_+= h_1+ C h_2
\ee
in terms of $h_{1,2}$ where
\be
C= \frac{\tilde M_{12}^2}{(1-b^2)k^2 + \tilde M_{11}^2 -\tilde M^2_{22}}.
\ee
Equivalently we have
\be
h_1= \frac{h_+ -C h_-}{1+C^2}, \ \ h_2=\frac{h_-+Ch_+}{1+C^2}
\ee
which will be useful when defining the Jordan frame graviton.
It is convenient  to define the characteristic wave number
\be
\bar k^2 = \frac{ \vert \tilde M_{11}^2 -\tilde M^2_{22}\vert }{\vert 1-b^2\vert}.
\ee
Hence
when $k\gg \bar k$, $C$ goes to zero in $1/k^2$ whilst when $k\ll \bar k$, $C$ goes to a constant or order one. In fact we have
\be
k\ll \bar k, \ \  C\sim \frac{\tilde M_{12}^2}{\tilde M_{11}^2 -\tilde M^2_{22}}
\ee
and
\be
k\gg \bar k, \ \ C\sim \frac{\tilde M_{12}^2}{\tilde M_{11}^2 -\tilde M^2_{22}} \frac{\bar k^2}{ k^2}.
\ee
The wave number $\bar k$ depends on how small the deviation
\be
\frac{\vert \omega_+- \omega_-\vert}{k} \sim \vert b-1\vert
\ee
can be, i.e. how small $\vert b-1\vert$ is.

The initial conditions for $\bar h_{\alpha}$ are related to the waves obtained in General Relativity (as  the size of the regions where the waves are created is smaller than the cosmological horizon and their energy is very large compared to $H_0$) scaled by $\frac{\beta_\alpha b_\alpha^{1/2}}{\beta_1^2 +\beta_2^2} \frac{a_J^4}{a_{\rm GR}^3 \tilde a_J}\vert_0 $, see (\ref{ini}), i.e.
\be
\bar h^{\alpha}_0=\frac{\beta_\alpha b_\alpha^{1/2}}{\beta_1^2 +\beta_2^2} \frac{a_J^4}{a_{\rm GR}^3 \tilde a_J}\vert_0 h_{\rm GR}
\ee
where the first denominator comes from the rescaling between the cosmological and local, i.e. physical, Newton constant and the fiducial one in the action.
This follows from the calculation in section 3 of the wave form emitted from a local source. The local source generates the initial wave which then propagate far away in a plane wave approximation.
The resulting waves after emission are then simply
\be
\bar h_{1}= \left(\frac{\beta_1}{\beta_1^2 +\beta_2^2}(\frac{e^{i\omega_+t} +C^2 e^{i\omega_- t}}{1+C^2}) + \frac{\beta_2 C}{\beta_1^2 +\beta_2^2}(\frac{e^{i\omega_+t} -e^{i\omega_- t}}{1+C^2})\right ) e^{-i\vec k.\vec x}\frac{a_J^4}{\tilde a_J a_{\rm GR}^3}\vert_0 h_{\rm GR}
\ee
and
\be
\bar h_{2}= \left(\frac{\beta_2 b^{1/2}}{\beta_1^2 +\beta_2^2}(\frac{e^{i\omega_-t} +C^2 e^{i\omega_+ t}}{1+C^2}) - \frac{\beta_1 C}{\beta_1^2 +\beta_2^2}(\frac{e^{i\omega_-t} -e^{i\omega_+ t}}{1+C^2})\right ) e^{-i\vec k.\vec x} \frac{a_J^4}{\tilde a_J a_{\rm GR}^3}\vert_0  h_{\rm GR}
\ee
As a result we get for the Jordan frame gravitational wave
\be
\bar h_J= a_J h_J= \beta_1 \bar h_1 + \beta_2 b^{1/2} \bar h_2
\ee
the following
\be
\bar h_J=(\frac{(\beta_1 +\beta_2 b^{1/2} C)^2}{1+C^2} e^{i\omega_+t}+\frac{(\beta_2 b^{1/2} -\beta_1 C)^2}{1+C^2} e^{i\omega_-t}) e^{-i\vec k.\vec x} \frac{a_J^4}{\tilde a_J a_{\rm GR}^3}\vert_0\frac{h_{\rm GR}}{\beta_1^2 +\beta_2^2}
\ee
This is the wave-form emitted by a far-away source when the gravitational waves show a birefringent behaviour.

\subsection{The effective speed of gravitational waves}

When the $b$ is very close to one, the wave generated by a distant source $A_J= \Re(h_J)$ reads
\be
A_J=(\frac{(\beta_1 +\beta_2 C)^2}{1+C^2} \cos( {i\omega_+t}-i\vec k.\vec x)+\frac{(\beta_2 -\beta_1 C)^2}{1+C^2} \cos({i\omega_- t}-i\vec k.\vec x)) \frac{a_J^4}{\tilde a_J a^3_{\rm GR}}\frac{h_{\rm GR}}{\beta_1^2 +\beta_2^2}
\ee
where $\omega_+\sim \omega_-$. Defining $\omega= \frac{\omega_++\omega_-}{2}$ and $\Delta \omega= \omega_--\omega_+$, we have
\begin{eqnarray}
A_J&=& \frac{(\beta_1 +\beta_2b^{1/2} C)^2+(\beta_2b^{1/2} -\beta_1 C)^2}{1+C^2}\cos (\Delta \omega t)  [ \cos (\omega t -i\vec k.\vec x) +\nonumber \\ &&  \frac{(\beta_1 +\beta_2b^{1/2} C)^2-(\beta_2b^{1/2}-\beta_1 C)^2}{(\beta_1 +\beta_2 b^{1/2} C)^2+(\beta_2b^{1/2} -\beta_1 C)^2}\tan (\Delta \omega t)\sin(\omega t -i\vec k.\vec x)  ] \frac{a_J^4}{\tilde a_J a^3_{\rm GR}}\vert_0 \frac{h_{\rm GR}}{\beta_1^2 +\beta_2^2}.\nonumber \\
\end{eqnarray}
This represents wave beats compared to the usual wave front of GR. When the two eigenfrequencies satisfy $ \Delta \omega t \ll 1$, the wave form can be cast into a propagating wave with a time dependent phase shift
\be
A_J = A \cos (\omega t -i\vec k.\vec x -\delta)
\ee
where the amplitude is given by
\be
A=\frac{(\beta_1 +\beta_2 b^{1/2} C)^2+(\beta_2b^{1/2} -\beta_1 C)^2}{1+C^2}\cos (\Delta \omega t)\frac{a_J^4}{\tilde a_J a^3_{\rm GR}}\vert_0 \frac{h_{\rm GR}}{\beta_1^2 +\beta_2^2}
\ee
 with a small time dependence and a phase shift
 \be
\delta= \frac{(\beta_1 +\beta_2b^{1/2} C)^2-(\beta_2b^{1/2} -\beta_1 C)^2}{(\beta_1 +\beta_2b^{1/2} C)^2+(\beta_2 b^{1/2}-\beta_1 C)^2}\tan (\Delta \omega t).
\ee
The wave propagates with the energy
\be
\omega= c_T k + \frac{\tilde M_{11}^2}{4k} +\frac{\tilde M_{22}^2}{4bk}
\ee
in an expansion in $\frac{\tilde M^2_{ij}}{k^2}$ and $k=\sqrt{\vec k^2}$.
We have introduced the effective speed of the gravitational waves
\be
c_T= \frac{1+b}{2}.
\ee
This effective speed is highly constrained when $b<1$, i.e. when the effective speed is less than the speed of light. Indeed in this case,  high energy cosmic rays can emit gravitons in a Cerenkov fashion and this would deplete the number count of cosmic rays on earth. This is not the case if and only if \cite{Moore:2001bv,Kimura:2011qn}
\be
(1-b) \lesssim 10^{-17}.
\ee
One can check check that in this case $\Delta \omega d\ll 1$ for sources such that $d\ll 100$ Mpc.
Of course the energy of cosmic rays is higher than the cut-off scale of doubly coupled bigravity so this constraint may be relaxed when considering the UV completion of the model.

As a result, we see that the effective speed of gravitational waves is extremely constrained by observations. In the following section, we will consider the prospects of detecting deviations from GR when the parameter $b$ is so tightly bounded.

\subsection{The emitted spectrum and detection prospects}

Let us now consider the spectrum of gravitational waves at a distance $d$ from the source. For that, it is convenient to consider the spectrum as obtained from the square of the amplitude
\be
\vert \bar h_J\vert^2 = (\frac{a_J^4}{\tilde a_J a_{\rm GR}^3}\vert_0)^2 (\frac{\beta_1^2 + b\beta_2^2}{\beta_1^2+\beta_2^2})^2(1-4 \frac{(\beta_1+\beta_2 b^{1/2}C)^2(\beta_2b^{1/2} -\beta_1 C)^2}{(\beta_1^2 +b\beta_2^2)^2}\sin^2 (\frac{(\omega_+-\omega_-)t}{2}))\vert h_{\rm GR}\vert^2.
\ee
This means that the signal has a change of amplitude and a time modulation, and that at a time $t=d$ 
\be
P_J(k) = (\frac{a_J^4}{\tilde a_J a_{\rm GR}^3}\vert_0)^2 (\frac{\beta_1^2 + b\beta_2^2}{\beta_1^2+\beta_2^2})^2(1-4 \frac{(\beta_1+\beta_2 b^{1/2}C)^2(\beta_2b^{1/2} -\beta_1 C)^2}{(\beta_1^2 +b\beta_2^2)^2}\sin^2 (\frac{(\omega_+-\omega_-)d}{2}))P_{\rm GR} (k)
\ee
the spectrum is modulated by a frequency dependent pre-factor. Let us first connect with the case of singly coupled gravity.
When $\beta_1$ or $\beta_2$ vanishes we find that
\be
\vert \bar h_J\vert^2 = (\frac{ a_J^4}{\tilde a_J a_{\rm GR}^3}\vert_0)^2 (1-4 C^2 \sin^2 (\frac{(\omega_+-\omega_-)d}{2}))\vert h_{\rm GR}\vert^2
\ee
where $C$ is constant for $k\lesssim \bar k$ and vanishes at large $k$ \cite{Narikawa:2014fua}. This retrieves the known results of the singly coupled case. In the doubly coupled case, the term in $\sin^2$ never vanishes, i.e.  this is a clear difference with the singly coupled case.

Let us notice that the modulation should only be effective when the variation of the $\sin^2$ term is not too rapid compared to the frequency of the signal in GR. If this is the case and the averaged $<\sin ^2>=1/2$ is used, the effects of bigravity are only a change in the amplitude of the signal, i.e. degenerate with the astrophysical features of the emitting system. On the other hand when
\be
\vert b-1 \vert \lesssim \frac{1}{k_{\rm exp} d}
\ee
where $k_{\rm exp}$ is the most sensitive frequency of the detecting device, and $d$ the distance to the emitting source, the modulation of the GR signal would be relevant. For sources around $d=100$ Mpc and a sensitivity peaking in the nano Hertz regime \cite{Manchester:2010tp}, we can
hope to observe effects for $\vert b-1\vert \lesssim 10^{-7}$, four orders of magnitude lower than the pulsar bound. The pulsar bound would be probed only by the detection of nano Hertz events in our immediate vicinity around $10$ kpc.

\section{Discussion and Conclusion}

We have discussed the emission and the propagation of gravitational waves in doubly coupled bigravity. The deviations from GR are essentially governed by one parameter $b$ which differs from one
only in the transient cosmological era between the matter era in the past and the future dark energy dominated one. It turns out that the deviation of this parameter from one measures the effective speed
of gravitational waves in bigravity. This can be constrained by both the absence of gravitational Cerenkov effect and the energy loss of binary pulsars. As a result, we do not expect that the effective speed of gravitational waves
differs from one by more than one per mil. This is still large enough to induce possible modulations of the wave form of the gravitational wave signal in the Jordan frame, i.e. the gravitational wave coupled to matter. The best prospect of detecting this gravitational birefringence would be with nano-Hertz interferometry experiments, and deviations of the gravitational speed up to $10^{-7}$ would be observable from sources further than 100 Mpc. Another way of detecting these effects would be to monitor nearby sources of both gravitational and electromagnetic waves and trying to detect a phase difference between these signals \cite{Bettoni:2016mij}.

The bound on $\vert b-1\vert \lesssim 10^{-3}$ from binary pulsar constraints implies that other effects of bigravity such as a change in the growth of cosmological structures would also be tightly restricted. Indeed, as an order of magnitude, the growth parameters such as $\mu$ and $\Sigma$ deviate from GR as $\vert b-1\vert$ and therefore one would not expect effects on structure formation much larger than the percent level. This would have implications for the detection of bigravity effect by future cosmological surveys \cite{Sawicki:2016klv}. The details of this comparison are left for future work.
\vskip 0.5 cm
\noindent {\bf Acknowledgements:} We would like to thank Juan Garcia-Bellido and Diego Blas for comments on the manuscript. We are grateful to Nicola Tamanini for remarks and discussions. This project has received funding from the European Union’s Horizon 2020 research and innovation programme under the Marie Skłodowska-Curie grant agreement No 690575, ACD acknowledges partial support from STFC under grants ST/L000385/1 and ST/L000636/1, JN acknowledges support from the Royal Commission for the Exhibition of 1851 and BIPAC. This article is based upon work related to the COST Action CA15117 (CANTATA) supported by COST (European Cooperation in Science and Technology).

\bibliographystyle{JHEP}
\bibliography{BGwave}

\end{document}